\documentclass[12pt,preprint]{emulateapj}
\usepackage{graphicx}
\usepackage{wasysym}

\shorttitle{Frequency of Potential Venus-Analogs}
\shortauthors{Stephen R. Kane et al.}
\slugcomment{Submitted for publication in the Astrophysical Journal
  Letters}

\begin{document}

\title{On the Frequency of Potential Venus Analogs from Kepler Data}
\author{
  Stephen R. Kane\altaffilmark{1},
  Ravi Kumar Kopparapu\altaffilmark{2,3,4,5,6},
  Shawn D. Domagal-Goldman\altaffilmark{7}
}
\email{skane@sfsu.edu}
\altaffiltext{1}{Department of Physics \& Astronomy, San Francisco
  State University, 1600 Holloway Avenue, San Francisco, CA 94132,
  USA}
\altaffiltext{2}{Department of Geosciences, Penn State University, 443
  Deike Building, University Park, PA 16802, USA}
\altaffiltext{3}{NASA Astrobiology Institute's Virtual Planetary
  Laboratory, P.O. Box 351580, Seattle, WA 98195, USA}
\altaffiltext{4}{Penn State Astrobiology Research Center, 2217 Earth
  and Engineering Sciences Building University Park, PA 16802, USA}
\altaffiltext{5}{Center for Exoplanets \& Habitable Worlds, The
  Pennsylvania State University, University Park, PA 16802, USA}
\altaffiltext{6}{Blue Marble Space Institute of Science, PO Box 85561,
  Seattle, WA 98145-1561, USA}
\altaffiltext{7}{NASA Goddard Space Flight Center, Greenbelt, MD
  20771, USA}


\begin{abstract}

The field of exoplanetary science has seen a dramatic improvement in
sensitivity to terrestrial planets over recent years. Such discoveries
have been a key feature of results from the {\it Kepler} mission which
utilizes the transit method to determine the size of the planet. These
discoveries have resulted in a corresponding interest in the topic of
the Habitable Zone (HZ) and the search for potential Earth
analogs. Within the Solar System, there is a clear dichotomy between
Venus and Earth in terms of atmospheric evolution, likely the result
of the large difference ($\sim$ factor of two) in incident flux from
the Sun. Since Venus is 95\% of the Earth's radius in size, it is
impossible to distinguish between these two planets based only on
size. In this paper we discuss planetary insolation in the context of
atmospheric erosion and runaway greenhouse limits for planets similar
to Venus. We define a ``Venus Zone'' (VZ) in which the planet is more
likely to be a Venus analog rather than an Earth analog.  We identify
43 potential Venus analogs with an occurrence rate ($\eta_{\venus}$)
of $0.32^{+0.05}_{-0.07}$ and $0.45^{+0.06}_{-0.09}$ for M dwarfs and
GK dwarfs respectively.

\end{abstract}



\section{Introduction}
\label{intro}

The sensitivity of exoplanet detection instrumentation and techniques
has dramatically improved over the past decade. This has enabled
significant progress towards the detection of terrestrial-sized
exoplanets using the radial velocity and transit methods. In
particular, the {\it Kepler} mission has yielded a plethora of
exoplanet candidates \citep{bor11a,bor11b,bat13,bur14}, many of which
have been confirmed by virtue of their multiplicity
\citep{lis14,row14}. The primary purpose of the {\it Kepler} mission
is to determine the frequency of Earth-sized planets in the Habitable
Zone (HZ), which is the region around a star where water can exist in
a liquid state on the surface of a planet with sufficient atmospheric
pressure. Several Earth-sized planets have been located in this region
as a result of these {\it Kepler} observations \citep{bor12,qui14}.

The inner and outer boundaries of the HZ for various main sequence
stars have been estimated using one-dimensional climate models by
\citet{kas93}. These boundary estimates have been revised in recent
years by the same group, including the extension to later spectral
types \citep{kop13a} and the consideration of different planetary
masses \citep{kop14}. The HZ calculations are available through the
Habitable Zone Gallery \citep{kan12a}, which provides HZ calculations
for all known exoplanetary systems. An important aspect of these HZ
calculations is that they provide a means to estimate the fraction of
stars with Earth-size planets in the HZ, or $\eta_{\oplus}$. Much of
the recent calculations of $\eta_{\oplus}$ utilize {\it Kepler}
results since these provide a large sample of terrestrial size objects
from which to perform meaningful statistical analyses
\citep{dre13,kop13b,pet13,tra12}.

The transit method has a dramatic bias towards the detection of
planets which are closer to the host star than farther away
\citep{kan08}. For example, the geometric transit probabilities of
Mercury, Venus, and Earth for an external observer are 1.2\%, 0.64\%,
and 0.46\% respectively. Additionally, a shorter orbital period will
result in an increased signal-to-noise (S/N) of the transit signature
due to the increased number of transits observed within a given
timeframe. The consequence of this is that {\it Kepler} has
preferentially detected planets interior to the HZ which are therefore
more likely to be potential Venus analogs than Earth analogs. An
example of this is Kepler-69c which was initially thought to be a
strong HZ candidate \citep{bar13} but subsequent analysis showed that
it's more likely to be a super-Venus \citep{kan13}. Since the
divergence of the Earth/Venusian atmospheric evolutions is a critical
component for understanding Earth's habitability, the frequency of
Venus analogs ($\eta_{\venus}$) is also important to quantify.

\begin{figure*}
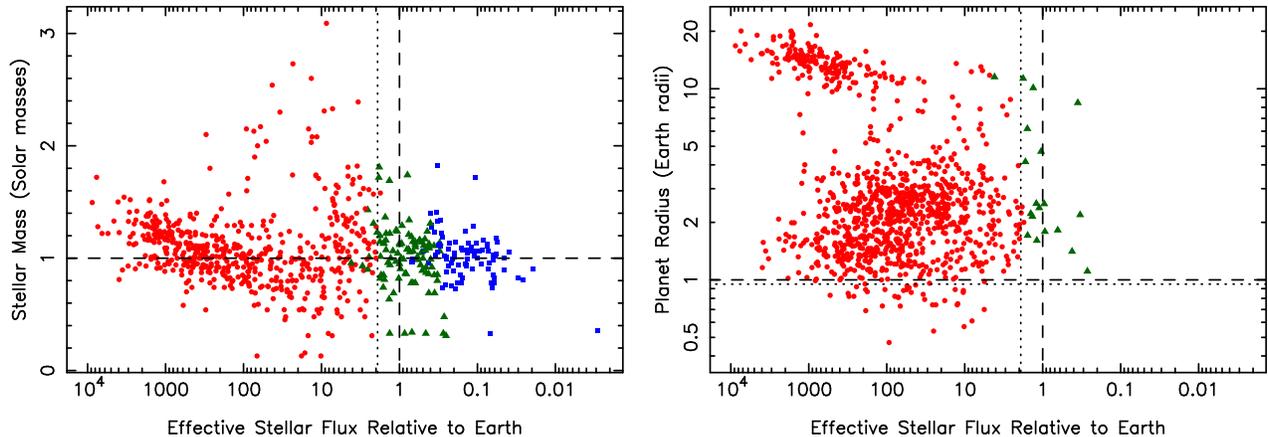

  \begin{center}
    \begin{tabular}{cc}
      \includegraphics[angle=270,width=8.2cm]{f01a.ps} &
      \includegraphics[angle=270,width=8.2cm]{f01b.ps}
    \end{tabular}
  \end{center}
  \caption{The dependence of effective stellar flux received by the
    planet on stellar mass (left) and planet radius (right). The green
    triangles are those planets which spend more than 50\% of their
    orbital phase within the optimistic Habitable Zone. The red
    circles and blue squares are planets interior and exterior to the
    Habitable Zone respectively. The dashed crosshairs show the
    location of Earth, and the dotted crosshairs show the location of
    Venus. The right-hand plot includes only those planets for which a
    radius determination is available.}
  \label{insolfig}
\end{figure*}

Here we provide a definition for the Venus Zone (VZ) and an estimate
for $\eta_{\venus}$ by examining the frequency of Venus analogs from
{\it Kepler} data. In Section \ref{insoldist} we discuss the current
insolation distribution of exoplanets based upon confirmed exoplanet
discoveries. In Section \ref{bound} we describe the outer boundary of
the VZ in terms of the runaway greenhouse limit (Section \ref{outer})
and the inner VZ boundary using an estimate of the atmospheric erosion
limit (Section \ref{inner}). In Section \ref{freq} we present
calculations of $\eta_{\venus}$ using exoplanet candidates from the
{\it Kepler} mission. Finally, we provide concluding remarks in
Section \ref{conclusion}.


\section{The Planetary Insolation Distribution}
\label{insoldist}

Exoplanets have been discovered at a variety of distances around
different kinds of stars. However, even without direct knowledge of
planetary surface conditions, one can still calculate the HZ of a star
based on Earth models. For example, the models of \citet{kas93} and
\citet{kop13a,kop14} quantify incident fluxes that would cause
planetary temperature conditions to sway to either a runaway
greenhouse effect or to a runaway snowball effect. For a given planet,
these calculations are based upon the stellar effective temperature,
stellar flux, and water absorption by the atmosphere. The flux
thresholds at which these transitions from one planetary state to
another occur allow for both conservative and optimistic scenarios
depending on how long it is presumed that Venus and Mars were able to
retain liquid water on their surfaces. Here we adopt the
``conservative'' and ``optimistic'' models of \cite{kan13}. The
conservative model uses the ``Runaway Greenhouse'' and ``Maximum
Greenhouse'' criteria for the inner and outer HZ boundaries
respectively. The optimistic model uses the ``Recent Venus'' and
``Early Mars'' criteria for the inner/outer HZ boundaries. These
criteria are described in detail by \citet{kop13a,kop14}.

The insolation (stellar flux received relative to Earth) distribution
for exoplanets are summarized in Figure \ref{insolfig}. The required
stellar and planetary data for confirmed exoplanets were extracted
from the Exoplanet Data Explorer\footnote{\tt http://exoplanets.org/}
\citep{wri11}. The data are current as of 21st June 2014. In each
plot, the red circles represent those planets interior to the
optimistic HZ and the blue squares those planets exterior to the
HZ. The green triangles represent those planets which spend more the
50\% of their orbital phase within the optimistic HZ, keeping in mind
that some of these planets lie in eccentric orbits. The dashed
crosshairs indicate Earth's position in the plots for comparison.

The stellar mass vs insolation plot (Figure \ref{insolfig}, left)
appears to contain a trend from high flux at high stellar mass to low
flux at low stellar mass. This is due to observational bias since
low-mass planets at longer orbital periods fall under current
detection thresholds. Figure \ref{insolfig} (right) includes only
those planets for which a radius determination is available. The
distinct giant and terrestrial planet populations at high incident
fluxes is made clear in this figure, which is likely a result of
planet formation/migration mechanisms. More importantly, this figure
is the key for determining $\eta_{\oplus}$ which can be thought of as
the relative number of points near the cross-hairs (Earth). As this
figure continues to fill out, it will be the clearest indicator of the
prevalence of Earth-analogs.

Another compelling question that arises from the plots in Figure
\ref{insolfig} is: how many of the points represent potential Venus
analogs as that is clearly where detection is currently more
efficient? In other words, missions such as {\it Kepler} are more
capable of determining $\eta_{\venus}$ than $\eta_{\oplus}$. Since
Venus and Earth are approximately the same size, the Venus analogs
will emerge earlier than the Earth analogs due to the increased
geometric transit probability \citep{kan08}. The location of Venus on
the plots is indicated by dotted crosshairs. The solar flux received
by the Earth is approximately 1365~W/m$^2$ and the flux received by
Venus is 2611~W/m$^2$, thus Venus receives 1.91 times more flux than
the Earth.  In order to determine $\eta_{\venus}$, we must first
define boundaries for the Venus Zone.


\section{Defining the Venus Zone Boundaries}
\label{bound}

Here we describe approximate boundaries for the Venus Zone (VZ) which
will be subsequently be used to estimate $\eta_{\venus}$.


\subsection{Outer Venus Boundary: Runaway Greenhouse}
\label{outer}

The outer boundary of the VZ is defined by the limit where oceans will
completely evaporate, leading to (for example) the apparent
Venus/Earth dichotomy that we see within our Solar System. The precise
mechanisms through which this takes place are complicated, but
essentially the loss of liquid water results in the inability to
execute a carbon cycle and thus efficiently moderate the levels of
atmospheric carbon (CO$_2$). There are various ways in which surface
temperature conditions might rise to such levels that would result in
a runaway greenhouse. \citet{barn13} suggested that tidal effects may
result in Venus analogs through tidal heating, particularly for
planets in eccentric orbits \citep{kan12b,wil02}. Here we consider
only the effects of stellar flux on the potential for runaway
greenhouse, as described by \citet{kas93} and \citet{kop13a,kop14}.

There are also a variety of ways in which the intrinsic properties of
the planet will influence this boundary. The calculations of
\citet{kop13a} do not include H$_2$O and CO$_2$ clouds, each of which
have separate effects. H$_2$O clouds generally move the inner HZ
boundary closer to the star \citep{lec13a,wol14,yan13} but CO$_2$
clouds result in IR back-scattering thus increasing the surface
temperature and moving the HZ outward through greenhouse
warming. Models by \citet{for97} and \citet{sel07} show that the
overall effect of neglecting clouds may result in an underestimate of
the greenhouse effect. \citet{abe11} and \citet{lec13b} describe how a
planet may lose surface water and remain habitable as a ``land
planet''. \citet{arm14} further showed that certain ranges of
obliquity will expand the HZ through seasonal variations, whilst
\citet{yan14} demonstrated the dependence of the HZ inner edge on
planetary rotation. However, the boundary at which water evaporation
and runaway greenhouse will take hold is well approximated by the
1~M$_\oplus$ Runaway Greenhouse of \citet{kop14}. We thus adopt this
as the outer boundary of the VZ.


\subsection{Inner Venus Boundary: Atmospheric Erosion}
\label{inner}

\begin{figure}
  \includegraphics[angle=270,width=8.2cm]{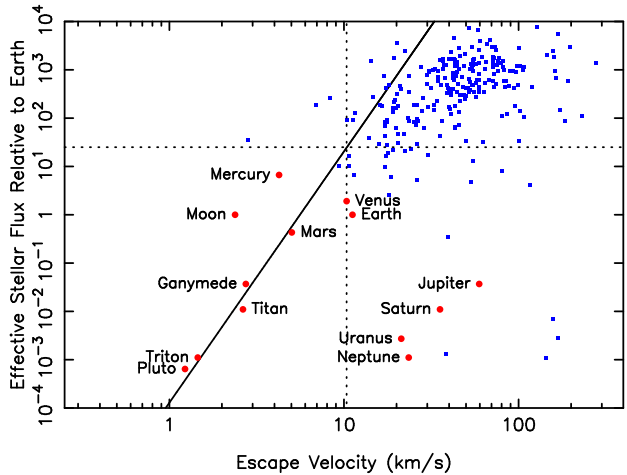}
  \caption{Stellar flux (relative to Earth) received by various
    objects as a function of their escape velocities. Selected objects
    from the Solar System are shown as labeled red circles. The small
    blue squares represent exoplanets with sufficient planetary and
    stellar information. The dotted corsshairs show the incident flux
    needed for Venus to be at the ``cosmic shoreline''.}
  \label{innerfig}
\end{figure}

Runaway greenhouse may be prevented if significant atmospheric mass
loss occurs as a result of proximity to the host star. The role of
extreme ultraviolet (XUV) radiation in atmospheric mass loss has been
the subject of several studies \citep{lam09,lop12,lop13}. The XUV
effect is applicable to terrestrial planets in high incident flux
regimes (less than 10 day orbital periods). Here we adopt a simple
approximation for the loss of an atmosphere based upon the work of
\citet{cat09,cat13} and referred to as the ``cosmic shoreline'' by
\citet{zah13}. This is an empirical relationship between incident flux
and surface gravity derived from observations of objects within the
Solar System and then extrapolated to exoplanets. Figure
\ref{innerfig} shows a new version of this relationship where selected
Solar System objects are shown as red circles. Confirmed exoplanets
for which sufficient data were available (see Section \ref{insoldist})
are shown as blue squares. The solid line is a power-law approximation
which separates those bodies with (right of the line) and without
(left of the line) atmospheres. Many of the exoplanets that lie to the
left of the line (blue squares) have high densities ($>5.5$~g/cm$^3$),
which if correct, would be consistent with iron-silicate
composition. We find that the best approximation for the power-law is
$I \propto v_e^{5.2}$ where $I$ is the incident flux and $v_e$ is the
escape velocity. There are several caveats to note about this
relationship between incident flux and atmospheric mass loss. There is
an implicit assumption regarding the mean molecular weight of the
atmospheric material which is being averaged to account for all
atmospheric compositions. Furthermore, the timescale for atmospheric
loss will vary depending on stellar mass and luminosity for a given
planet.  With these caveats in mind, we adopt the incident flux needed
for Venus to be at the ``cosmic shoreline'' as an approximation for
the inner boundary of the VZ; $\sim$~25 times the Earth incident flux
(see Figure \ref{innerfig}).


\section{The Frequency of Venus Analogs}
\label{freq}

\begin{figure*}
  \begin{center}
    \begin{tabular}{cc}
      \includegraphics[width=8.5cm]{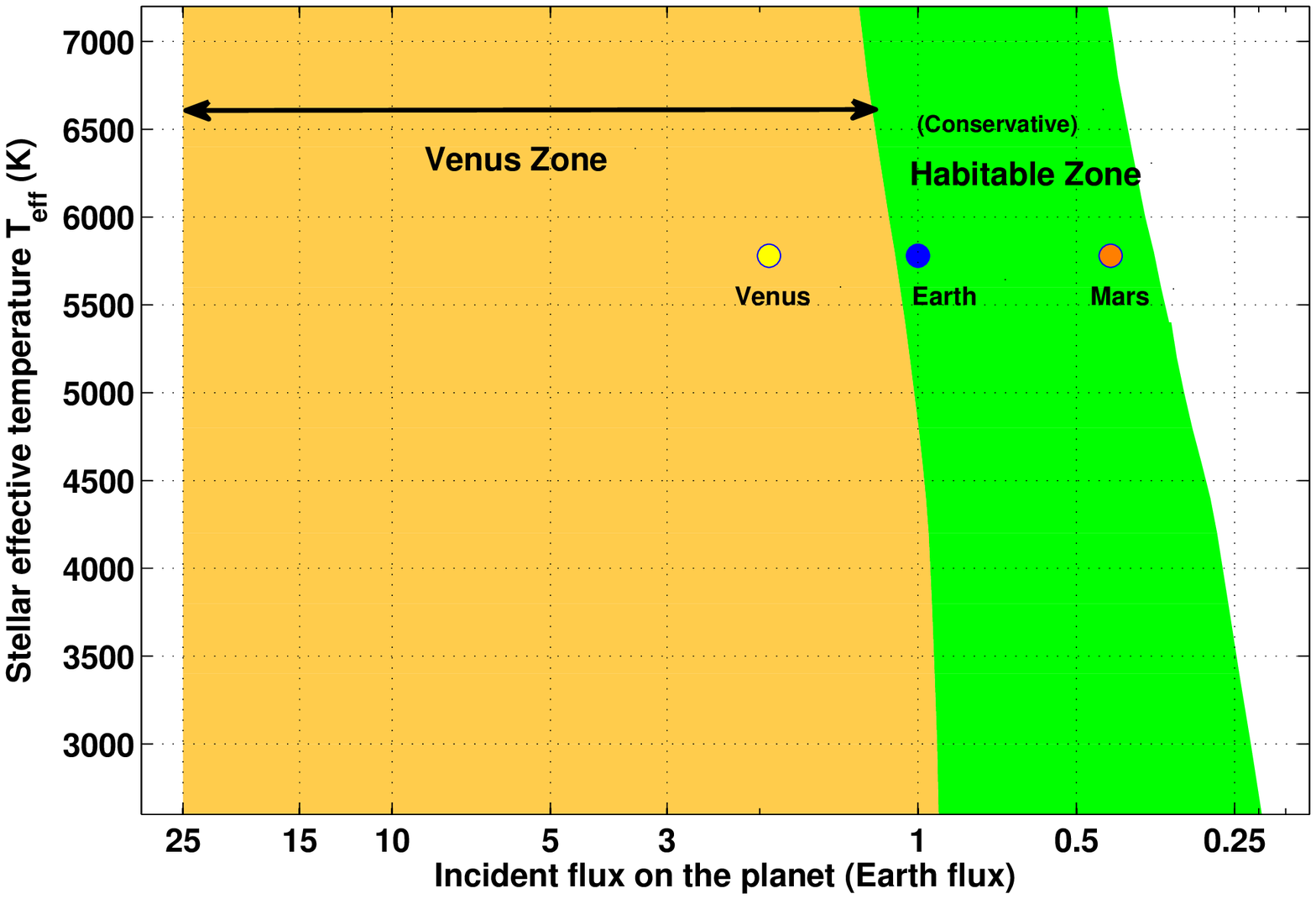} &
      \includegraphics[width=8.2cm]{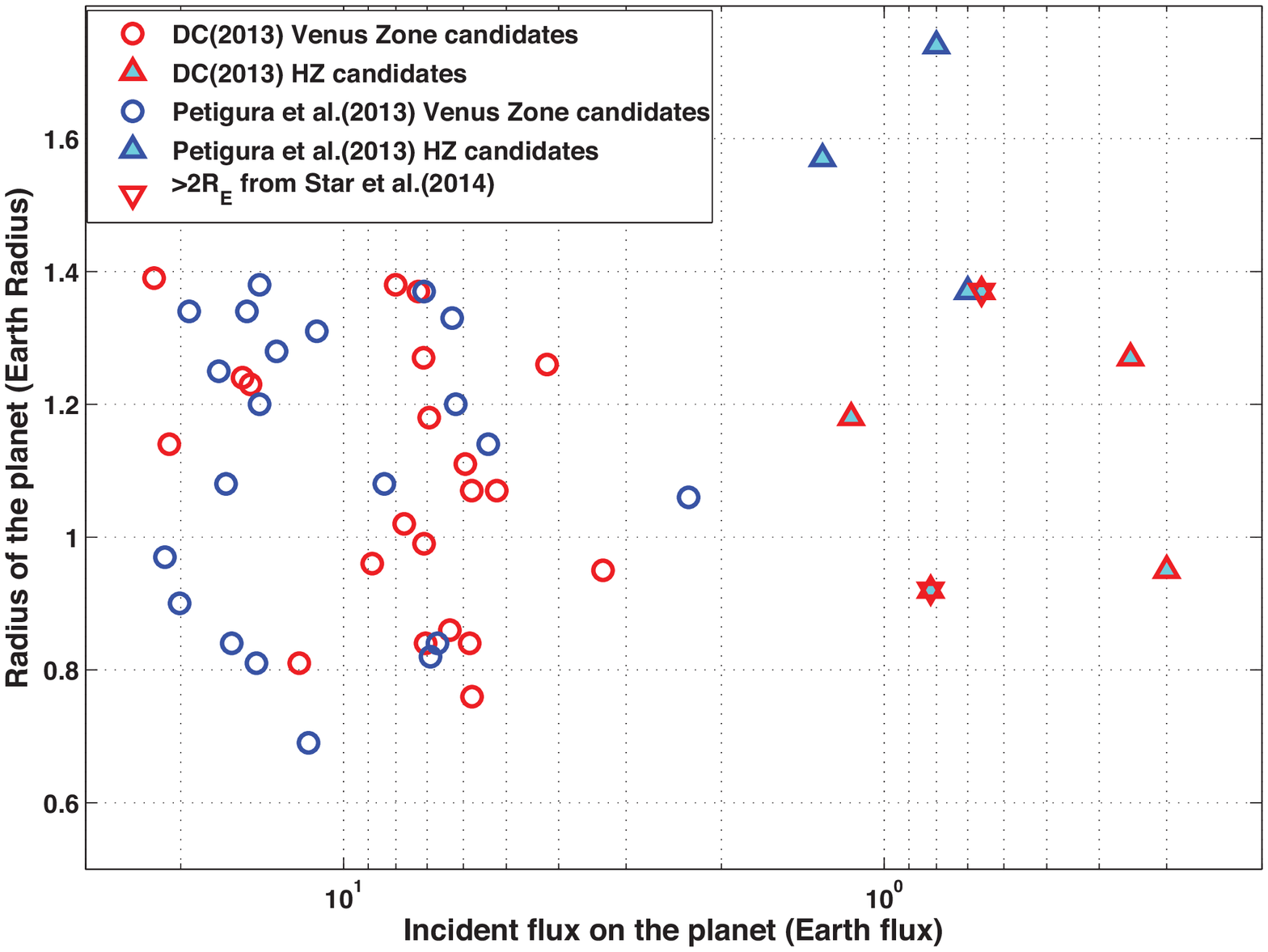}
    \end{tabular}
  \end{center}
  \caption{Left: Incident stellar flux on a planet versus stellar
    effective temperature, showing the extent of the Venus Zone and
    (conservative) Habitable Zone. Right: Incident stellar flux on a
    planet versus planet radius. The data are from \citet{dre13} (red
    circles and triangles) and \citet{pet13} (blue circles and
    triangles). Planets in the HZ from both the data sets are in
    filled colors. There are a total of $43$ planets in the VZ. List
    of these candidates are provided in Tables \ref{table1} \&
    \ref{table2}. Two of the planets from \citet{dre13} (red down
    triangles), KOI 1422.02 and KOI 2626.01, may be larger than
    $2$R$_{\oplus}$, according to \citet{sta14}.}
  \label{koihz}
\end{figure*}

In this section, we will calculate the occurrence rate of terrestrial
size planets that receive a flux which puts them between the runaway
greenhouse limit and atmospheric erosion caused by incident stellar
flux. We call this region the ``Venus Zone'' (VZ) to identify those
exoplanets that are either in the runaway greenhouse state or in the
process of losing their atmosphere due to the close proximity to their
host star (see Section \ref{bound}). The extent of the VZ and HZ
boundaries are shown in the left panel of Figure \ref{koihz} as a
function of stellar effective temperature. We calculate the occurence
rate of VZ planets for GKM stellar spectral types, using the data from
\citet{dre13} (for M-dwarfs) and \citet{pet13} (for K \&
G-dwarfs). Both these studies calculate the planet occurrence rate
based on the following equation:
\begin{eqnarray}
  f(R_{p},P) &=& \sum_{i=1}^{N_{p}(R_{p},P)} \frac{a_{i}}{R_{\star,i}
    N_{\star,i}}
  \label{rateeq}
\end{eqnarray}
where $a_{i}$ is the semi-major axis of planet $i$, $R_{\star,i}$ is
the host star's radius of planet $i$, $N_{\star,i}$ is the number of
stars around which planet $i$ could have been detected and
$N_{p}(R_{p},P)$ is the number of planets with the radius $R_{p}$ and
period $P$. The ratio $a_{i}/R_{\star,i}$ is the inverse of the
probability of transit orientation, which is considered to take
non-transiting geometries into the estimation of occurrence
rate. \citet{pet13} use $N_{\star,i} = C \times n_{\star}$, where $C$
is the completeness correction factor for each $(P,R_{p})$ bin, and
$n_{\star} = 42,557$ stars in their survey of brightest (Kepler
magnitude 10--15) GK spectral types.

To calculate the occurrence rate for M-dwarfs, we use the
$N_{\star,i}$ values provided within the period-radii cells of Figure
15 of \citet{dre13}, which give the number of stars around which a
planet from the center of the grid cell would have been detected with
a signal-to-noise ratio above $7.1 \sigma$. This should still give us
nearly the same occurrence rate, or an underestimate of the actual
value. This is because, as pointed by \citet{kop13b}, the
$N_{\star,i}$ values calculated by \citet{dre13} and used in their
estimate of $\eta_{\oplus}$ are generally lower in value than the
center of the grid cells values we use from Figure 15. If this is the
general trend, i.e., if we are overestimating $N_{\star,i}$
systematically, then our calculated occurrence rates for M-dwarfs can
be considered as a lower bound to the actual value.

In the case of occurrence rates for K \& G-dwarfs, completeness
correction factors from \citet{pet13} are available (see Figure~ S11,
supplementary information). As mentioned above, they also provide
$n_{\star} = 42,557$ stars. Therefore, all the required values are
available to calculate the occurrence rate using Equation
\ref{rateeq}.

We identify those planets that are in the appropriate period-radius
bin from each data set. We consider planets with $0.5-1.4$R$_{\oplus}$
as terrestrial size, and count all the planets in the VZ. Restricting
the outer edge of VZ to the runaway greenhouse flux limit from
\citet{kop14} for different stars, and the inner edge to the empirical
relation between stellar flux and planetary gravity for Venus from
Figure \ref{innerfig} ($25$ times incident flux on Earth), there are
22 planets in the M-star sample of \citet{dre13}. With these
conditions, using Equation \ref{rateeq}, the occurrence rate of
terrestrial size planets that are potentially Venus analogs around
M-dwarfs is $0.32^{+0.05}_{-0.07}$.  Similarly, there are $21$ planets
in the K $\&$ G-star sample of \citet{pet13} within the VZ. The
corresponding occurrence rate around K $\&$ G-stars is
$0.45^{+0.06}_{-0.09}$. The uncertainties are estimated by (1)
calculating the effective number of stars searched (detected planets
in the sample divided by the occurrence rate) (2) constructing a
binomial cumulative distribution function (CDF) representing the
likelihood of finding specified number of planets given the effective
number of stars searched and the probability of finding a planet (i.e,
occurrence rate), (3) finding the number of planets corresponding to
the $15.9$ and $84.1$ percentiles in the CDF, (4) calculating the new
occurrence rates by dividing the number of planets from percentile
ranges with the effective number of stars searched and (5) finally
subtracting these new rate bounds from the original rates. A list of
all the planet candidates are provided in Tables \ref{table1} $\&$
\ref{table2}. A comparison of the various classifications of the
candidates as described here is shown in the right panel of Figure
\ref{koihz}.

Several caveats are to be noted: (1) None of these are confirmed
planets, and are only candidates. In fact, a recent study by
\citet{sta14} showed that two of the planet candidates (KOI~1422.02
and KOI~2626.01) in the HZ are larger than previous estimates. This
may remove some planets out of contention and thus reduce the
occurrence rate. The uncertainties in the stellar parameters and thus
planetary radii are not well constrained, so this may also add to the
uncertainty in our occurrence estimate \citep{kan14}. On the other
hand, \citet{dre13} used the Q1--Q6 sample. It is likely that an
expanded dataset may increase the occurrence rate around M-dwarf
stars. So the net effect may not change the rate estimate
significantly. (2) We use an upper flux limit of $25$ times Earth
flux. This number arises from Figure \ref{innerfig}, by finding the
intersection point for Venus on the ``cosmic shoreline'' (the solid
line). Our terrestrial size planet has a radius range of $0.5-1.4$
R$_{\oplus}$. Correspondingly, the flux limit can vary depending on
the size (or gravity) of the planet. This in turn changes the
occurrence rate (i.e, a planet with lower gravity has a lower flux
limit, and hence a lower estimate of the occurrence rate). Our
estimates above are in this sense an average value that encompasses
the terrestrial size planets. (3) Although we define an outer boundary
of the VZ, there are planetary properties that could push this out
even further, such as sufficiently increased CO$_2$ levels that could
push an atmosphere into runaway greenhouse even at larger
distances. However, the probability of runaway greenhouse will start
to drop dramatically with increasing distance beyond the VZ. Thus, our
calculation of $\eta_{\venus}$ using the defined VZ likely encompasses
most Venus analogs. (4) Just as planets in the HZ should be considered
``Earth-like candidates'' until their properties can be confirmed via
spectroscopy, planets inside the VZ should be considered ``Venus-like
candidates'' until spectral measurements and characterization efforts
are undertaken.

\begin{deluxetable*}{lcccc}
  \tablecolumns{5}
  \tablewidth{0pc}
  \tablecaption{\label{table1} List of M-dwarf KOIs that are
    terrestrial in size ($0.5-1.4$ R$_{\oplus}$) and which receive a
    stellar flux between runaway greenhouse limit and atmospheric
    erosion. Data from \citet{dre13}.}
  \tablehead{
    \colhead{KOI} &
    \colhead{Radius ($R_{\oplus}$)} &
    \colhead{Period (days)} &
    \colhead{$T_\mathrm{eff}$ (K)} &
    \colhead{Flux ($F_{\oplus}$)}
  }
  \startdata
$251.02$&$   0.76^{+0.08}_{-0.10}$&$  5.78$& $3743.0$& $16.81$\\
$ 571.01$&  $1.37^{+0.14}_{-0.21}$&$  7.27$& $3820.0$& $13.98$\\   
$ 571.04$&  $1.39^{+0.14}_{-0.21}$&$ 22.41$& $3820.0$& $ 3.12$\\
$ 886.01$&  $1.38^{+0.30}_{-0.27}$&$  8.01$& $3579.0$& $ 5.30$\\ 
$ 886.02$&  $0.81^{+0.18}_{-0.16}$&$ 12.07$& $3579.0$& $ 3.07$\\   
$ 886.03$&  $1.14^{+0.25}_{-0.22}$&$ 21.00$& $3579.0$& $ 1.47$\\  
$ 899.01$&  $1.27^{+0.15}_{-0.25}$&$  7.11$& $3587.0$& $ 8.74$\\ 
$ 899.02$&  $0.95^{+0.12}_{-0.19}$&$  3.31$& $3587.0$& $24.26$\\
$ 899.03$&  $1.24^{+0.15}_{-0.24}$&$ 15.37$& $3587.0$& $ 3.13$\\   
$1085.01$&  $1.02^{+0.11}_{-0.10}$&$  7.72$& $3878.0$& $14.89$\\  
$1146.01$&  $0.99^{+0.13}_{-0.10}$&$  7.10$& $3778.0$& $12.32$\\ 
$1422.01$&  $0.84^{+0.19}_{-0.19}$&$  5.84$& $3424.0$& $ 4.20$\\
$1681.01$&  $1.18^{+0.18}_{-0.15}$&$  6.94$& $3608.0$& $ 8.63$\\   
$1843.01$&  $1.26^{+0.14}_{-0.22}$&$  4.20$& $3584.0$& $19.30$\\  
$1843.02$&  $0.86^{+0.10}_{-0.15}$&$  6.36$& $3584.0$& $11.09$\\ 
$1867.03$&  $1.07^{+0.11}_{-0.12}$&$  5.21$& $3799.0$& $20.76$\\
$2036.02$&  $1.07^{+0.10}_{-0.20}$&$  5.79$& $3903.0$& $21.85$\\   
$2057.01$&  $1.11^{+0.10}_{-0.14}$&$  5.95$& $3900.0$& $21.67$\\  
$2179.01$&  $1.23^{+0.15}_{-0.18}$&$ 14.87$& $3591.0$& $ 3.20$\\ 
$2191.01$&  $0.96^{+0.11}_{-0.13}$&$  8.85$& $3724.0$& $ 8.61$\\
$2650.01$&  $1.18^{+0.40}_{-0.15}$&$ 34.99$& $3735.0$& $ 1.15$\\
$2650.02$&  $0.84^{+0.29}_{-0.11}$&$  7.05$& $3735.0$& $ 9.73$
  \enddata
\end{deluxetable*}

\begin{deluxetable*}{llcccc}
  \tablecolumns{6}
  \tablewidth{0pc}
  \tablecaption{\label{table2} List of K $\&$ G-dwarf KOIs that are
    terrestrial in size ($0.5-1.4$ R$_{\oplus}$) and which receive a
    stellar flux between runaway greenhouse limit and atmospheric
    erosion. Data from \citet{pet13}.}
  \tablehead{
    \colhead{KIC ID} &
    \colhead{KOI} &
    \colhead{Radius ($R_{\oplus}$)} &
    \colhead{Period (days)} &
    \colhead{$T_\mathrm{eff}$ (K)} &
    \colhead{Flux ($F_{\oplus}$)}
  }
  \startdata
$1849702 $ &  $2538.01 $ &$1.37 $&$39.83 $&$ 5023.0 $&$ 7.1$\\
$ 3223433$ &  $ 4548.01$ &$ 1.20$&$ 61.08$&$  5518.0$&$  6.2$\\
$ 4276716$ &  $ 1619.01$ &$ 0.81$&$ 20.66$&$  4882.0$&$ 14.5$\\
$ 5374403$ &  $ 2556.01$ &$ 1.38$&$ 40.84$&$  5523.0$&$ 14.3$\\
$ 5706966$ &  $ 1908.01$ &$ 1.28$&$ 12.55$&$  4350.0$&$ 13.3$\\
$ 5780930$ &  $ 3412.01$ &$ 1.20$&$ 16.75$&$  4683.0$&$ 14.3$\\
$ 7742408$ &  $    0.00$ &$ 0.84$&$ 33.50$&$  4802.0$&$  6.7$\\
$ 8349399$ &  $ 4763.01$ &$ 1.25$&$ 56.45$&$  5980.0$&$ 17.0$\\
$ 8481129$ &  $ 2402.01$ &$ 1.08$&$ 16.30$&$  4763.0$&$ 16.5$\\
$ 8560940$ &  $ 3450.01$ &$ 0.97$&$ 31.97$&$  5852.0$&$ 21.4$\\
$ 8625732$ &  $ 4701.01$ &$ 0.84$&$ 31.97$&$  5597.0$&$ 16.1$\\
$ 9116075$ &  $    0.00$ &$ 0.82$&$ 24.50$&$  4570.0$&$  6.9$\\
$ 9150827$ &  $ 1408.01$ &$ 1.08$&$ 14.53$&$  4252.0$&$  8.4$\\
$ 9334893$ &  $ 2298.01$ &$ 0.90$&$ 16.67$&$  4922.0$&$ 20.1$\\
$ 9412760$ &  $ 1977.01$ &$ 1.34$&$  9.39$&$  4346.0$&$ 19.3$\\
$ 9718066$ &  $ 2287.01$ &$ 0.69$&$ 16.09$&$  4470.0$&$ 11.6$\\
$ 9993683$ &  $    0.00$ &$ 1.34$&$ 29.94$&$  5241.0$&$ 15.1$\\
$10453588$ &  $ 2484.01$ &$ 1.31$&$ 68.89$&$  5739.0$&$ 11.2$\\
$11462341$ &  $2124.01 $ &$1.06 $&$42.34 $&$ 4252.0 $&$ 2.3$\\
$11774991$ &  $2173.01 $ &$1.14 $&$37.81 $&$ 4705.0 $&$ 5.4$\\
$12735830$ &  $ 3311.01$ &$ 1.33$&$ 31.83$&$  4712.0$&$  6.3$
  \enddata
\end{deluxetable*}


\section{Conclusions}
\label{conclusion}

A critical question that exoplanet searches are attempting to answer
is: how common are the various elements that we find within our own
Solar System? This includes the determination of Jupiter analogs since
the giant planet has undoubtedly played a significant role in the
formation and evolution of our Solar System. When considering the
terrestrial planets, the attention often turns to atmospheric
composition and prospects of habitability. In this context, the size
degeneracy of Earth with its sister planet Venus cannot be ignored and
the incident flux must be carefully considered. Here we have provided
a definition for a region where terrestrial planets are potentially
Venus analogs using data from Solar System bodies. This results in the
first estimate of $\eta_{\venus}$; the frequency of Venus/Earth-size
planets within the VZ. The occurrence rates are similar to those of
$\eta_\oplus$ though with significantly smaller uncertainties since
{\it Kepler} is more biased towards planets in the VZ compared with
the HZ. Future missions capable of identifying key atmospheric
abundances for terrestrial planets will face the challenge of
distinguishing between possible Venus and Earth-like surface
conditions. Discerning the actual occurrence of Venus analogs will
help us to decode why the atmosphere of Venus so radically diverged
from its sister planet, Earth.


\section*{Acknowledgements}

We thank the anonymous referee for helpful comments which improved the
manuscript. The authors would like to thank Eric Lopez and Kevin
Zahnle for several useful discussions. The authors would also like to
thank Courtney Dressing for her input in deriving uncertainties on
occurrence rates. R.K gratefully acknowledges funding from NASA
Astrobiology Institute's Virtual Planetary Laboratory lead team,
supported by NASA under cooperative agreement NNH05ZDA001C. This work
has made use of the Habitable Zone Gallery at hzgallery.org.


\end{document}